\documentclass{article}
\usepackage{graphicx} 
\usepackage{amsmath}
\usepackage{hyperref}
\usepackage{authblk}
\usepackage{tcolorbox}
\usepackage{enumitem}
\usepackage[numbers]{natbib}

\title{Fees in AMMs: \\A quantitative study}

\author{Abe Alexander\thanks{abealexander@outlook.com} \and Lars Fritz\thanks{lsfritz@proton.me}}

\begin{document}

\maketitle
\begin{abstract}
In the ever evolving landscape of decentralized finance automated market makers (AMMs) play a key role: they provide a market place for trading assets in a decentralized manner. For so-called bluechip pairs, arbitrage activity provides a major part of the revenue generation of AMMs but also a major source of loss due to the so-called 'informed orderflow'. Finding ways to minimize those losses while still keeping uninformed trading activity alive is a major problem in the field. In this paper we will investigate the mechanics of said arbitrage and try to understand how AMMs can maximize the revenue creation or in other words minimize the losses. To that end, we model the dynamics of arbitrage activity for a concrete implementation of a pool and study its sensitivity to the choice of fee aiming to maximize the revenue for the AMM. We identify dynamical fees that mimic the directionality of the price due to asymmetric fee choices as a promising avenue to mitigate losses to toxic flow. This work is based on and extends a recent article by some of the authors (\cite{abelars}).
\end{abstract}
\section{Introduction}

Decentralized Finance (DeFi) has revolutionized the cryptocurrency space by providing a broad range of financial services without the need for intermediaries. A key strategy in DeFi is arbitrage, where traders take advantage of price discrepancies for assets across various platforms or exchanges.

Arbitrage in DeFi functions similarly to traditional finance but benefits from the decentralized nature of blockchain technology. Traders can capitalize on price differentials between decentralized exchanges (DEXs), lending protocols, liquidity pools, and other DeFi platforms. These discrepancies arise due to differences in supply and demand, transaction delays, or pricing algorithm inefficiencies.

Arbitrageurs (ARBs) are vital for maintaining market efficiency within DeFi. By correcting price discrepancies, they help align prices across platforms and stabilize the market. However, arbitrage opportunities are typically short-lived as they attract other traders, leading to rapid price convergence.

Despite its profitability, DeFi arbitrage involves risks such as impermanent loss, transaction fees, and smart contract vulnerabilities. Additionally, regulatory uncertainties and protocol risks add complexity to arbitrage strategies in DeFi.

In summary, arbitrage in DeFi is a dynamic and lucrative aspect of decentralized finance, allowing traders to profit from price differentials across various protocols. It requires a deep understanding of market dynamics, smart contracts, and risk management. As DeFi evolves, arbitrage will continue to be a fundamental strategy shaping the decentralized finance landscape.

In the context of automated market makers (AMMs), arbitrage significantly influences revenue generation but also contributes to losses due to 'informed orderflow'. Examining various AMMs reveals that most fee generation comes from 'bluechip' pairs prevalent in both DeFi and centralized exchanges (CEXs). It is estimated that over $80\%$ of trading activity in these pairs involves arbitrage deals.

In this paper, we model the trading activity of an AMM assuming that all trading is arbitrage-driven. We assume that arbitrage happens between the AMM and an infinite liquidity source, which could be a centralized exchange (CEX). We model the price discovery of said CEX by means of geometric Brownian motion with drift, without additional assumptions. Simultaneously, we simulate an AMM, choosing a constant function type of pool. The results can easily be extended to more complex AMMs, such as concentrated liquidity implementations like Uniswap v3 pools.

\begin{figure}
\begin{center}
\includegraphics[width=0.85\textwidth]{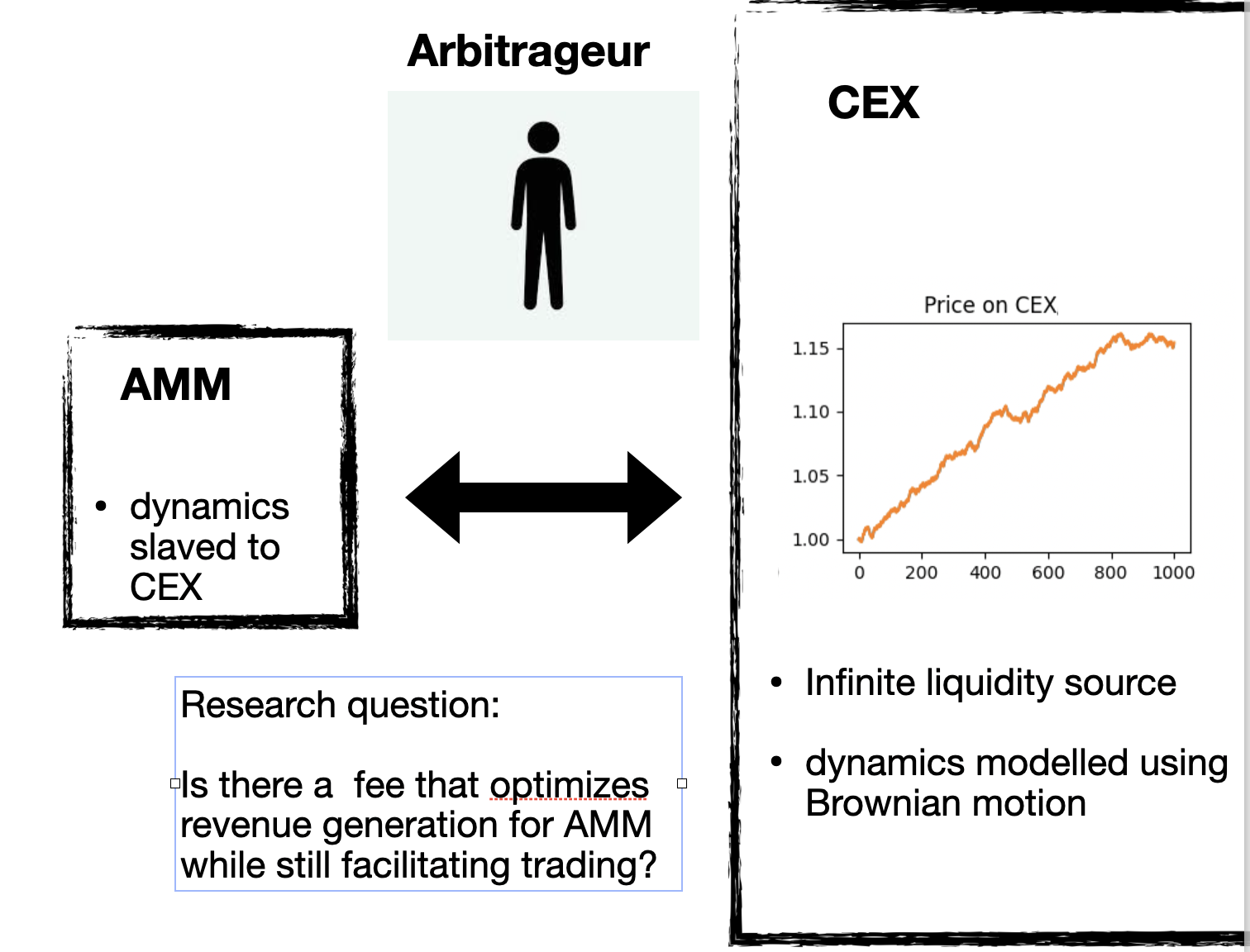}{\caption{The question studied revolves around the interplay between an AMM and a CEX and focuses on arbitrage. The main research question is whether there is an ideal fee setting that best allows to mitigate the losses due to toxic orderflow.}}\label{figintro}
\end{center}
\end{figure}

The focus of this paper is to gain a better understanding of the interplay between fee choice of the AMM and the ensuing arbitrage dynamics. The driving question is whether there are 'ideal' fee setting that facilitate arbitrage activity but at the same time minimize losses/maximize revenue due to toxic flow see Fig.~\ref{figintro}. 
Our study suggests that dynamical directional fees provide the sweetspot approach to optimize the revenue from arbitrage activities.

We attempted to keep the paper self-contained and to require a minimum of additional reading. It is organized as follows. We first review properties of the random walk and geometric Brownian motion in Sec.~\ref{sec:technical}. We proceed to introduce the mathematical description of the AMM as well as the analysis of an optimized arbitrage cycle in Sec.~\ref{sec:setup}. In Sec.~\ref{sec:simulations} we compare two different arbitrage strategies and find that the results agree quantitatively and qualitatively. We furthermore discuss how the findings can be rationalized using principles of the random walk introduced in Sec.~\ref{sec:technical}. We end with conclusions and an outlook on possible follow-up questions and implementations in Sec.~\ref{sec:conclusion}.

\noindent{\bf{Related literature:}}

 AMMs can be traced back to \cite{hanson2007logarithmic} and \cite{othman2013practical} with early implementations discussed in \cite{lehar2021decentralized}, \cite{capponi2021adoption}, and \cite{hasbrouck2022need}. Details of implementation are described in \cite{Adams20} and \cite{Adams21} as well as in a very recent textbook \cite{ottina2023automated}.

We study ways to optimize fees based on an arbitrage-only assumption. Uniswap v3 (\cite{Adams21}) addresses this problem by letting liquidity providers choose between different static fee tiers. Other automated market makers have implemented dynamic fees on individual pools, including Trader Joe v2.1 (\cite{mountainfarmer22joe}), Curve v2 (\cite{egorov21curvev2}) and Mooniswap (\cite{bukov20mooniswap}), Algebra (\cite{Volosnikov}), as well as \cite{Nezlobin2023}. Some of the general properties of toxic flow and loss versus rebalancing have been discussed in Refs.~\cite{Faycal1,Faycal2,Faycal3,milionis2024automated,crapis2023optimal,angeris2024multidimensional}. The paper is an extension of a recent article~\cite{abelars}.

\section{Random walks and their connection to market dynamics}\label{sec:technical}

Quite literally, a random walk constitutes a path of successive random steps. It is a paradigm in probability theory and has applications in various fields such as physics, computer science, biology, and economic theory. 

It is defined as a sequence of random steps on some mathematical space such as the integers, a lattice, or a graph. The simplest form of a random walk that also happens to be relevant to our purposes is the one-dimensional random walk, which can be described as follows:
We start the path at some point, usually the origin, \( x = 0 \) and then take a step left or right at each time step with probabilities $[p_{\rm{left}},p_{\rm{right}}]$ with $p_{\rm{left}}+p_{\rm{right}}=1$. In the following we will identify 'left' with 'up', 'right' with 'down', the position 'x' with the price of the asset 'p', and the origin with the starting price $p_0$.

Let \( p_n \) be the price after \( n \) steps. It is given by

\begin{eqnarray}\label{eq:randomwalk}
p_n = p_0+ p_0\sigma \sum_{i=1}^n \xi_i
\end{eqnarray}

where \( \xi_i \) are independent random variables taking values \( +1 \) or \( -1 \) with probability $p_{\rm{up}}$ and $p_{\rm{down}}$, respectively. We will see below that $\sigma$ plays the role of the standard deviation and will be related to the volatility when we discuss geometric Brownian motion.

Symmetric random walks, {\it {i.e.}} $p_{\rm{up}}=p_{\rm{down}}=1/2$, exhibit interesting properties some of which we review here: The average price after $n$ steps is $\langle p \rangle=p_0$ and the variance is $\langle (p_n-p_0)^2   \rangle=\langle p_n^2\rangle -p_0^2=p_0^2\sigma^2 n$.

In this paper we use the random walk to model the price of the CEX. However, we will find that it is also the best way to rationalize the findings of the simulations. 

A crucial concept in our analysis is the 'hitting' time which we define and analyze below.

\subsection{The random walk and the hitting time}

\begin{figure}
\begin{center}
\includegraphics[width=0.8\textwidth]{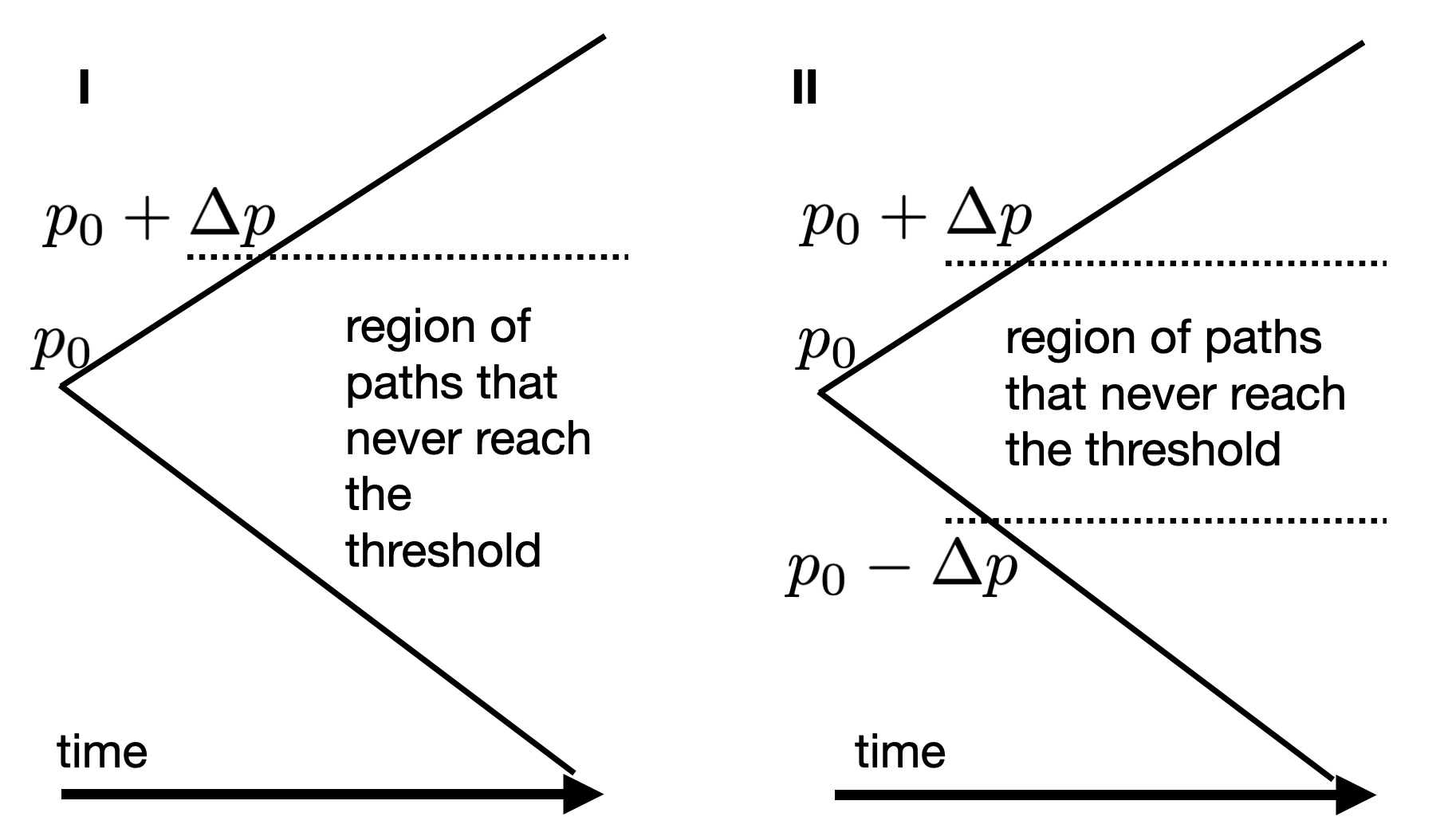}\caption{This figure shows how the space of reachable positions grows over time, shown by the opening wedge. I: in the case of a threshold only in upwards direction we observe that the number of paths that never reach the threshold grows at a comparable rate as the number of total paths. Consequently, the expectation value for the hitting time is infinite, as argued in the text. II: In the case of symmetric thresholds we find that the space of paths that never reach the threshold does not grow over time but is restricted to a ribbon, whereas the space of total paths does. These considerations assume a symmetric walk with $p_{\rm{up}}=p_{\rm{down}}$}\label{figure2}
\end{center}
\end{figure}

A question that will turn out as vital is how long or equivalently how many steps it takes on average to reach a specific price $p'=p_0+\Delta p$ when we start from $p_0$ and evolve according to the rules of Eq.~\eqref{eq:randomwalk}. It turns out, that this reveals one of the paradoxes of random walks. Counterintuitively, the answer is that the average number of steps it takes is infinite for a symmetric random walk. The reason for this is the following: there are many paths that never reach $p'=p+\Delta p$. In fact, any path that only explores prices below $p'$ qualifies for that, see Fig.~\ref{figure2} I. Over time, the fraction of the paths that never reach $p'$ grows at a comparable rate to the total number of paths. Consequently, the fraction of those paths compared to all the paths is not of measure zero implying that the average time it takes to reach $p_0+\Delta p$, henceforth referred to as 'hitting time', is formally infinite (we comment on this feature when we discuss the numerical simulations). It is also important to note that the hitting time becomes finite if the walk is not symmetric and there is a net directional movement in the direction of the threshold. In that case, the hitting time will scale linear with the threshold.

The more relevant question for arbitrage is not about a single threshold and its hitting time, but rather the hitting time for leaving a region bounded in upwards and downwards direction. The reason for that is that arbitrage activity is not restricted to price increases but also happens in the other direction, using the other asset. 
We find that there are two cases that can be distinguished and we start with the simpler one. Both of the cases will be important for understanding the findings of the arbitrage algorithm.
\subsubsection{Symmetric thresholds}
The underlying idea is that there is a minimum threshold price difference to be taken for arbitrage to be profitable (we discuss this explicitly in Sec.~\ref{sec:setup}). We assume that reaching $p_0\pm\Delta p$ gives access to profitable arbitrage activity. So it is a natural question to ask when this threshold can be reached on average.
The space of paths that never reach {\bf{any of the two points}} is much smaller compared to when we choose {\bf{just one threshold}}. In fact they are restricted to lie entirely inside a small ribbon, see Fig.~\ref{figure2} II. There are still many paths in this ribbon, but considerably less. Importantly, the ratio of those paths to the total number of paths goes to zero in the infinite time limit. This implies, that the hitting time is not infinite, but in fact finite.
This is backed up by simulations as discussed below. We distinguish two case, one in which the probability of going up and down is balanced, called 'symmetric walk', and another in which there is a preferential direction, referred to as 'asymmetric walk'.

{\bf{Symmetric walk: $p_{\rm{up}}=p_{\rm{down}}$}}

\begin{figure}
\begin{center}
\includegraphics[width=0.8\textwidth]{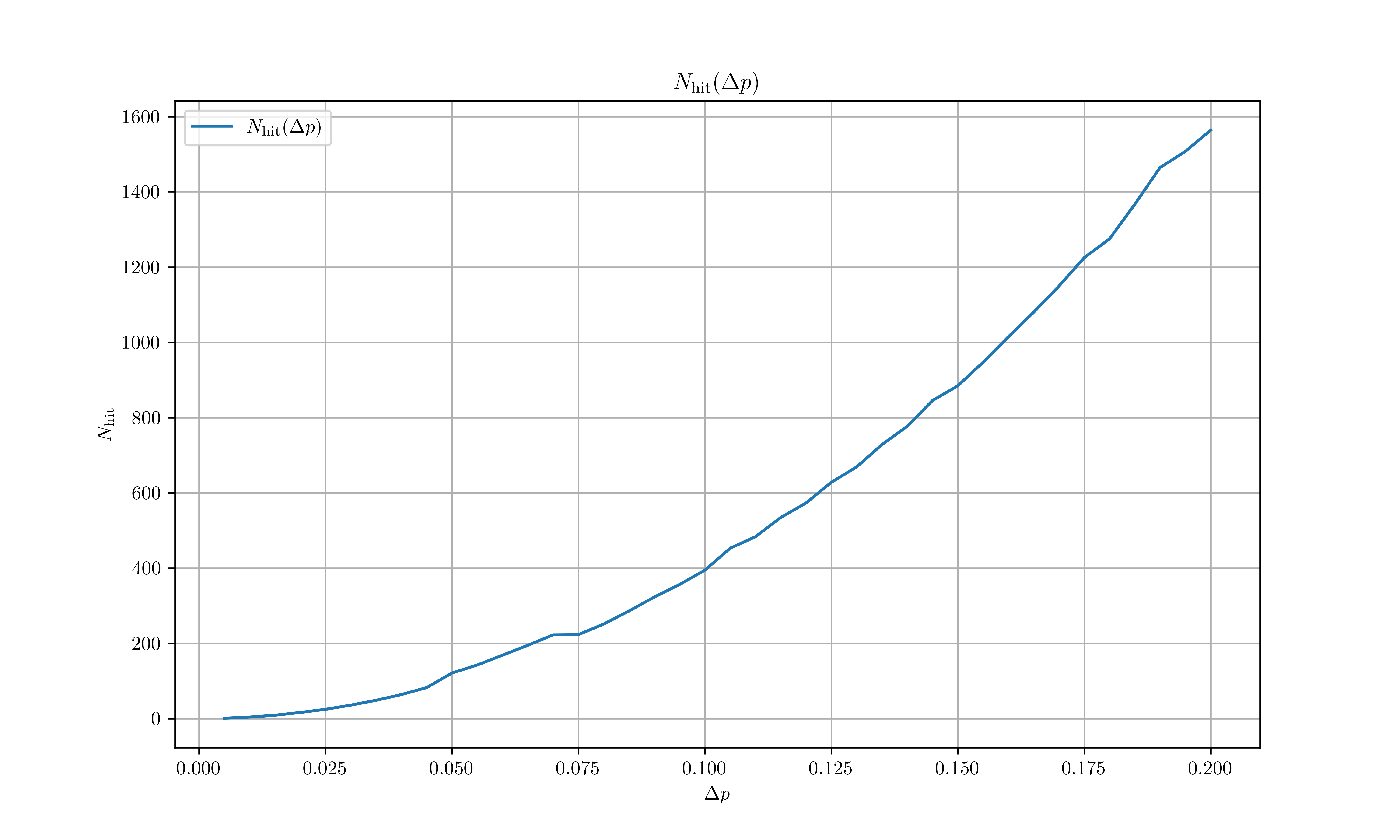}\caption{The hitting time for symmetric thresholds grows quadratic with the threshold.}\label{figure3}
\end{center}
\end{figure}

In this case, we made a simulation in the following way: in all cases we use $\sigma=0.02$ (which is a rather big value) and run $10000$ simulations. We only stop the individual run once the path 'hits' either $p_0+\Delta p$ or $p_0+\Delta p$ for a given $\Delta p$ (compare case II in Fig.~\ref{figure2}). For each run, we store the number of steps that were required to reach one of the two points and then average over all $10000$ runs which defines $N_{\rm{hit}}$ as the average hitting time (we drop average from now on). We repeat this as a function of $\Delta p$ and the result is shown in Fig.~\ref{figure3}. The important observation is that the hitting time scales quadratic with the threshold $\Delta p$, meaning $N_{\rm{hit}}\propto \Delta p^2$.

This can be rationalized starting from a generic property of a random walk that follwos the discussed rules. The mean displacement in a random walk scales like $\sqrt{\langle (p_n-p_0)^2   \rangle} =p_0\sigma \sqrt{n}$ where $n$ is the number of steps. We can assume that the mean displacement reaching the threshold value $\Delta p$ defines the hitting time $N_{\rm{hit}}$. This implies that we have to meet the condition $\Delta p=\sqrt{\langle (p_n-p_0)^2   \rangle} \propto \sqrt{N_{\rm{hit}}}$ or conversely $N_{\rm{hit}}(\Delta p)\propto \Delta p^2$, as e found numerically. (As a side note, we have run the simulation for only a one-sided threshold in which theory predicts an infinite hitting time. As expected, the simulation never converged and had to be stopped.)

{\bf{Asymmetric walk: $p_{\rm{up}}>p_{\rm{down}}$}}

\begin{figure}
\begin{center}
\includegraphics[width=0.8\textwidth]{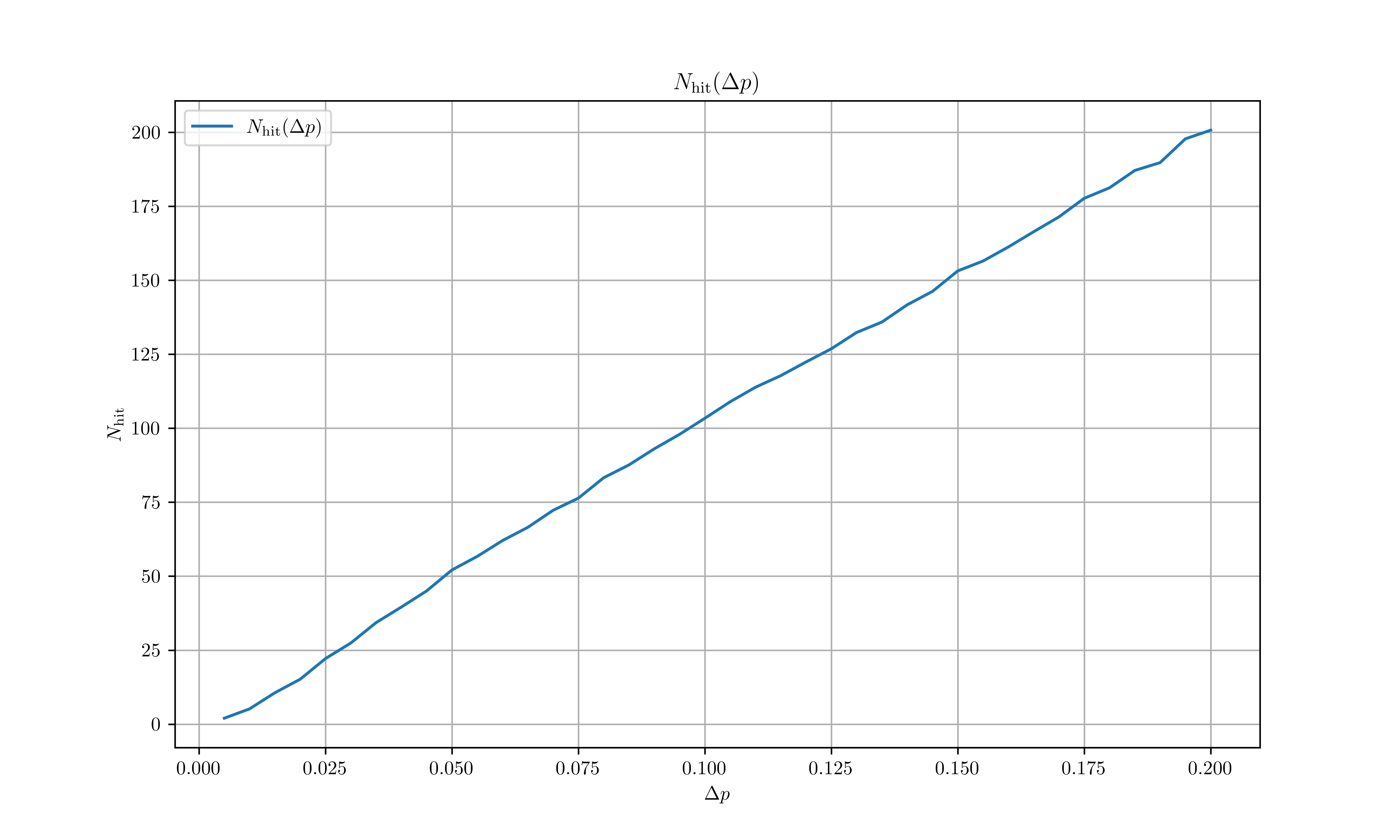}\caption{The hitting time for symmetric thresholds with drift grows linearly with the threshold.}\label{figure4}
\end{center}
\end{figure}

The situation with asymmetry can be analyzed in analogous manner, the only difference being that the wedges shown in Fig.~\ref{figure2} lose probability below $p_0$. There is an important modification, which is that the behavior of the hitting time goes over to linear behavior, see Fig.~\ref{figure4}. 
This can be related to the fact that with an asymmetry in movement, the average $\langle p_n \rangle-p_0 \propto n$. This implies that the average crosses the one of the two thresholds, depending on the direction of asymmetry, after a number of steps that grows linearly with the threshold. This implies $N_{\rm{hit}}\propto \Delta p$ as backed up by the numerics. As mentioned above, this property even persists with just a single threshold, Fig.~\ref{figure2} I, if the asymmetry leads to a mean displacement towards the threshold. In the opposite direction, the hitting time is still infinite.

\subsubsection{Asymmetric thresholds}

\begin{figure}
\begin{center}
\includegraphics[width=0.8\textwidth]{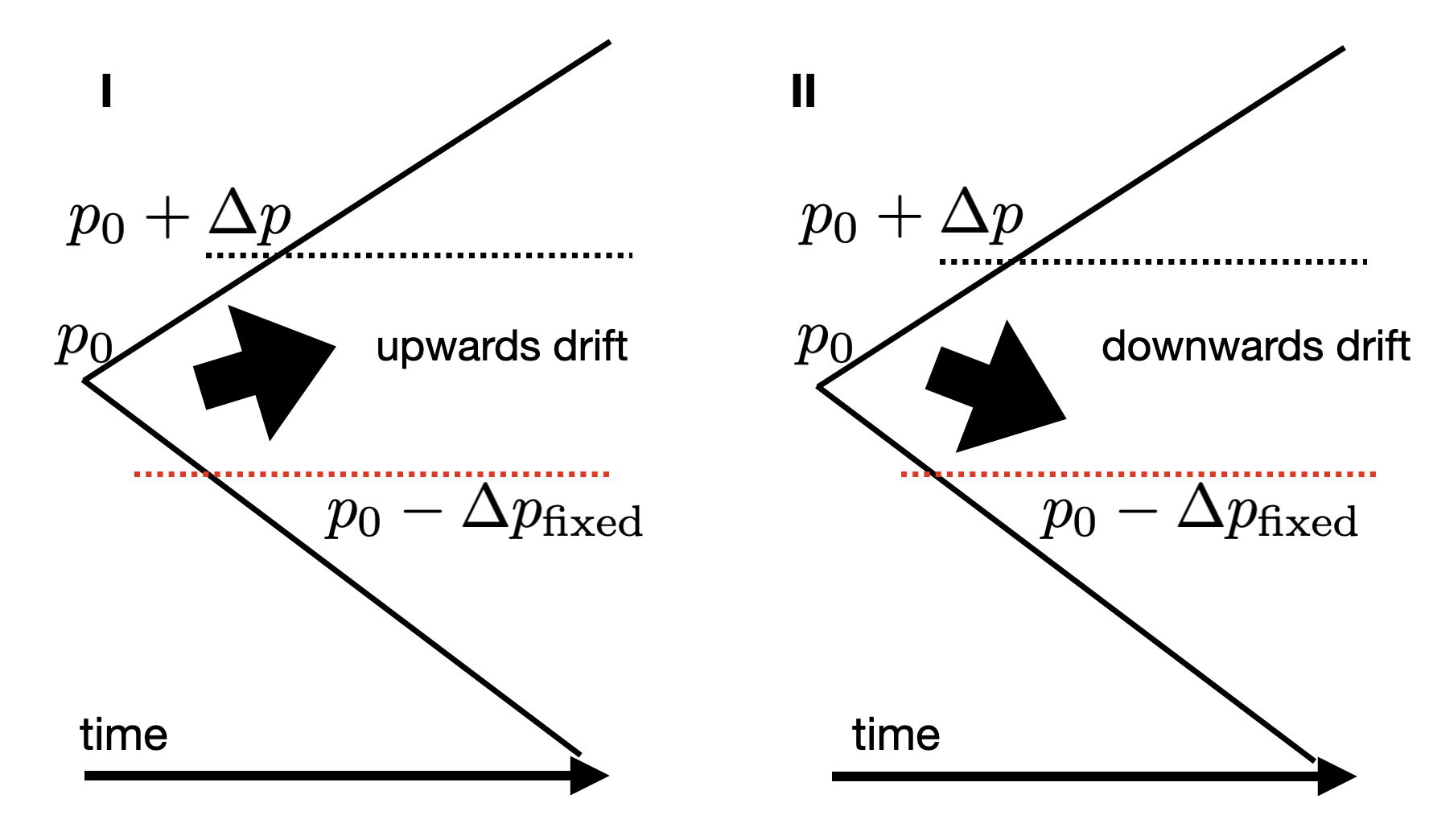}\caption{I: The upper threshold is varied whereas the lower kept fixed. The preferential direction is towards the upper threshold. II: Same situation as in I, only that the preferential direction is towards the lower threshold. }\label{figure5}
\end{center}
\end{figure}

It turns out, another relevant situation is to understand the case of asymmetric thresholds where one is fixed and the other one can move. We denote the threshold that is varied $\Delta p$ while the threshold that remains fixed and finite is called $\Delta p_{\rm{fixed}}$. The situation is depicted in Fig.~\ref{figure5}.

{\bf{Symmetric walk: $p_{\rm{up}}=p_{\rm{down}}$}}

\begin{figure}
\begin{center}
\includegraphics[width=0.99\textwidth]{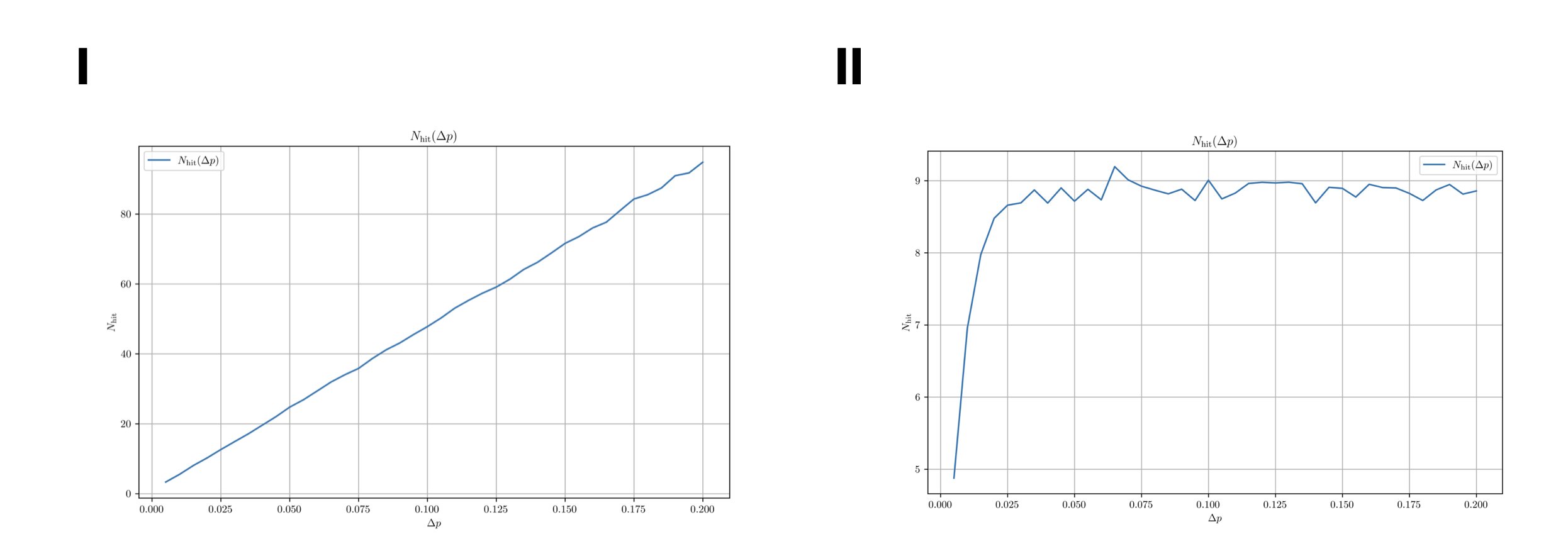}\caption{I: The hitting time grows linearly with the threshold for upwards movement whereas it saturates for downwards movement, II.}\label{figure6}
\end{center}
\end{figure}

We looked at the same situation in which both boundaries moved before and found that the hitting time increased quadratically with increasing the threshold. On the other hand, with one threshold being fixed at $\Delta p_{\rm{fixed}}$, we expect that in one direction there should be a hitting time that is independent of the upper threshold (not entirely true since in principle one expects infinite with just one). 
Interestingly, we find that the hitting time grows linearly with the threshold $\Delta p$ for a given fixed $\Delta p_{\rm{fixed}}$, see Fig.~\ref{figure5}. We are not including a plot of the result here since it looks identical to the one shown in Fig.~\ref{figure6} I, although the setup is without drift here.

{\bf{Asymmetric walk:}}

The expectation here is that if the preferential movement is towards the threshold that is varied, we should find the equivalent behavior, meaning a linear scaling of the hitting time with the threshold. On the other hand, if the preferential movement is directed towards the fixed threshold, the result should not depend on $\Delta p$ for $\Delta p>\Delta p_{\rm{fixed}}$.

The results are shown in Fig.~\ref{figure6} for the two situations side-by-side. In the case of the preferential movement being towards the direction in which the threshold is varied, situation I, we indeed find $N_{\rm{hit}}\propto \Delta p$ like in the previous section for the same reason: the average will cross the threshold after a number of steps that corresponds to the drift. In the other case, II, where we are moving towards the fixed threshold, the other moving threshold does not matter and we hit the threshold at a constant value, meaning we find $N_{\rm{hit}}=const$ for values of $\Delta p>\Delta p_{\rm{fixed}}$.   

\subsection{Geometric Brownian motion}

Geometric Brownian Motion (GBM) is a stochastic process commonly used to model the dynamics of financial assets. It is defined by the following stochastic differential equation (SDE):

\begin{equation}
    dp_t = \mu p_t \, dt + \sigma p_t \, dW_t,
    \label{eq:GBM}
\end{equation}

where \( p_t \) is the asset price at time \( t \), \( \mu \) is the drift coefficient, \( \sigma \) is the volatility coefficient, and \( W_t \) is a standard Brownian motion or Wiener process. In practice, it corresponds to a random variable $\pm1$ coming with equal probability.

The solution to the SDE in Equation \eqref{eq:GBM} can be obtained using Itô's Lemma, which provides a way to find the differential of a function of a stochastic process. Applying Itô's Lemma to \( p_t \), we get:

\begin{equation}
    d(\ln p_t) = \left( \frac{1}{p_t} dp_t - \frac{1}{2p_t^2} (dp_t)^2 \right).
\end{equation}

Substituting \( dp_t \) from Equation \eqref{eq:GBM}, we have:

\begin{equation}
    d(\ln p_t) = \left( \frac{\mu p_t \, dt + \sigma p_t \, dW_t}{p_t} - \frac{1}{2} \frac{\sigma^2 p_t^2 \, dt}{p_t^2} \right) = \left( \mu - \frac{\sigma^2}{2} \right) dt + \sigma dW_t.
\end{equation}

Integrating both sides from 0 to \( t \), we obtain:

\begin{equation}
    \ln p_t = \ln p_0 + \left( \mu - \frac{\sigma^2}{2} \right) t + \sigma W_t.
\end{equation}

Exponentiating both sides gives the explicit solution to the GBM SDE:

\begin{equation}
    p_t = p_0 \exp \left( \left( \mu - \frac{\sigma^2}{2} \right) t + \sigma W_t \right),
\end{equation}

where \( p_0 \) is the initial asset price at \( t = 0 \).

GBM is widely used in financial modeling due to its properties: the logarithm of the asset price follows a normal distribution, and it ensures that the asset price remains positive. These characteristics make it suitable for modeling stock prices and other financial instruments. In the following we will use GBM to model the CEX price of the asset.

\section{Modelling an AMM and an arbitrage cycle}\label{sec:setup}

We consider a simplified version of an arbitrage problem with two main players: an AMM and an infinite liquidity source, for instance embodying a CEX. We assume that price discovery happens on the CEX and the price dynamics of the AMM is 'slaved' to it. Furthermore, we assume that arbitrage can always be performed in an optimal fashion. We will discuss the implications of this below. 

\subsection{The AMM}

For simplicity, we model the AMM as a constant function market maker. The results, however, are more generic and in principle apply to any curve based form of AMM with a continuous pricing function. Examples include concentrated liquidity or other versions like Dodo's version of a pricing curve. We explain below how details enter the discussion. 

The basic equation that describes a constant function market maker is given by
\begin{eqnarray}\label{eq:cfmm}
x_{A}*x_B=L^2
\end{eqnarray}
where $x_A$ is the number of tokens of type $A$ and $x_B$ the number of tokens of type $B$. $L$ is conventionally called the liquidity and it is a measure for the price stability of the AMM. 

Swaps in AMMs are subject to price impact. There are, in principle, three prices worth discussing: the spot price, $p_s$, the effective swap price, $p_{\rm{eff}}$, and the spot price after a swap, $p_s'$, all of which are different. The price before a swap, the spot price, is simply defined by
\begin{eqnarray}
p_s=\frac{x_B}{x_A}\;.
\end{eqnarray}
It is the price that is charged for an infinitesimal swap with no price impact. The other prices, $p_{\rm{eff}}$ and $p_s'$ incorporate nonlinear effects due to the swap itself. 
To show this, let us consider a simple swap in which $\Delta A$ tokens are swapped for $\Delta B$ tokens. Using the constant function, Eq.~\eqref{eq:cfmm}, we find
\begin{eqnarray}
\Delta B= p_{\rm{eff}} \Delta A= \frac{p_s}{1+\frac{1}{x_A} \Delta A} \Delta A
\end{eqnarray}
where $1/x_{A}$ accounts for the non-linearity of the pricing function. This part will depend on the exact form of the AMM. For a generic AMM it would read
\begin{eqnarray}
\Delta B= p_{\rm{eff}} \Delta A= \frac{p_s}{1+p_{iA} \Delta A} \Delta A
\end{eqnarray}
where $p_{iA}=1/x_A$ is a general measure for the price impact since it accounts for the leading nonlinearity. Here, however, we continue with the special case of the CFMM. After the swap has been carried out, we end up with a new spot price given by
\begin{eqnarray}
p_s'=\frac{x_B-\Delta B}{x_A+\Delta A}=p_s \frac{1}{\left(1+\Delta A/x_A\right)^2}\;.
\end{eqnarray}
Consequently, for this sequence, we find that $p_s>p_{\rm{eff}}>p_s'$.
For the reverse swap, roles reverse and we have
\begin{eqnarray}
\Delta A= \frac{p^{-1}_s}{1+ \Delta B/x_B} \Delta B\;.
\end{eqnarray}

This constitutes the basis for the ensuing simulation of the dynamics of the AMM and is all that is required.

\subsection{The arbitrage cycle}

Now we have a look at the arbitrage cycle. An arbitrage agent observes that token A is cheaper inside the AMM than on the CEX. In concrete terms, this means that $p_{\rm{CEX}}/p_s>1$ where $p_{\rm{CEX}}$ is the price on the CEX. 
Consequently, it appears profitable to buy token A from the AMM and sell it on the CEX. The obvious question is: How can the arbitrage agent optimize the profit? To answer this, we need an equation that accounts for all gains and losses in the process. It is important here to reiterate that we assume that the CEX is of infinite liquidity and there is no price impact. While in practice the CEX is not of infinite liquidity, it is reasonable to assume that it is of significantly higher liquidity justifying the subsequent approach.

In the actual process, ARB takes a flashloan of size $\Delta FL$ (here we assume that it is taken in the denomination token B avoiding more swaps). For that, ARB has to  pay the fee$f_{\rm{fl}}$ meaning in total it costs $\Delta FL f_{fl}$. Additionally, the AMM charges a fee $f$ for swaps, meaning the actual $\Delta B$ entering the pool will be 
\begin{eqnarray}
\Delta B=\Delta FL(1-f)\;.
\end{eqnarray}
(Here we assume that the flashloan fee is payed back at the end of the cycle; later we will neglect it altogether)
After the swap to $\Delta A$ is completed, $\Delta A$ is sold at the CEX and ARB receives 
\begin{eqnarray}
\Delta B_{\rm{final}}=p_{CEX}\Delta A \;
\end{eqnarray}
at zero price impact.
Overall, this cycle results in a net profit
\begin{eqnarray}
P_B&=&p_{CEX}\Delta A-\Delta FL(1+f_{fl})-TXN \nonumber \\ &=& \frac{p_{CEX}}{p_s}\frac{\Delta FL(1-f)}{1+ \Delta FL(1-f)/x_B} -\Delta FL(1+f_{fl})-TXN
\end{eqnarray}
where $TXN$ accounts for the cost of transactions. In the following, we use the parameter $\alpha=p_{CEX}/p_s$ as a measure for the relative price of CEX and AMM. 

\subsubsection{Condition for successful arbitrage} 

We can start by asking when an arbitrage can be performed successfully. Clearly, the question is equivalent to asking when do we have $P_B>0$ as a function of $f$ and $\alpha$. To answer this, we first rewrite
\begin{eqnarray}
P_B=\Delta FL \left(\alpha \frac{1-f}{1+\Delta FL (1-f)/x_B}-(1+f_{fl})-\frac{TXN}{\Delta FL}) \right)\;.
\end{eqnarray}
Successful arbitrage in this direction can only happen for $\alpha>1$ (for $\alpha<1$ roles between $A$ and $B$ are reversed) but this is not sufficient. We are assuming for now that transaction costs are negligible ($TXN=0$), meaning we have a condition
\begin{eqnarray}
\alpha \frac{1-f}{1+p_{iA}\Delta FL (1-f)}-(1+f_{fl})>0\;.
\end{eqnarray}
We find that there is a threshold condition on $\alpha$ for a set AMM fee $f$ which needs to be overcome defined by
\begin{eqnarray}\label{eq:cond}
\alpha>\frac{1+f_{fl}}{1-f}\;.
\end{eqnarray}
In the remainder of this paper we will discuss $f_{\rm{fl}}=0$ meaning we find 
\begin{eqnarray}
\alpha>\alpha_{\rm{min}}=\frac{1}{1-f}
\end{eqnarray}
 as a condition.

There is an optimal $\Delta FL^{opt}$ from the point of view ARB that allows to maximize the gain $P^A$. This can be determined from 
\begin{eqnarray}
\frac{dP^A}{d \Delta FL}\Bigg|_{\Delta FL=\Delta FL^{\rm{opt}}}=0\;.
\end{eqnarray}

The solution is given by

\begin{eqnarray}
\Delta FL^{opt}=x_B \sqrt{\alpha_{\rm{min}}}\left(\sqrt{\alpha}-\sqrt{\alpha_{\rm{min}}}\right)
\end{eqnarray}
which is guaranteed to be positive once we fulfil Eq.~\eqref{eq:cond}. This amount does not align the price of the AMM with the CEX but slightly undershoots it. The price it ends up with is in fact given by $p_s'=p_{\rm{CEX}}(1-f)$ in the case of optimal arbitrage. This can easily be rationalized by dividing the arbitrage into 2 steps. The first cycle would end up at $p_s'=p_{\rm{CEX}}(1-f)$ and the second would not be profitable anymore so it would not be executed. It has one important consequence, though, which is any ensuing price movement on the CEX the increases the price immediately leads to an arbitrage cycle (this will proof important in the interpretation). 
The last thing to check is the corresponding optimal gain which is given by 

\begin{eqnarray}
P_A^{opt}=x_B\left(\sqrt{\alpha}-\sqrt{\alpha_{\rm{min}}}\right)^2\;
\end{eqnarray}
which is positive for situations that fulfil Eq.~\eqref{eq:cond}, as it should be.

\subsubsection{Optimal fees}

Arbitrage is an inevitable and most of the time even desired occurrence. Under stable conditions it is a valuable source of revenue for the AMM due to fee generation. In general, however, it leads to a loss for the AMM since overall it loses value in the process. One question is how to minimize this impact. It turns out that this is not necessarily easy to answer and there are many approaches to mitigate this loss. One of them is dynamical fees and we are investigating the optimal choice of fees for a given price differential $\alpha$. The situation is that there is a price difference between the CEX and the AMM with $alpha>1$. We are now looking for the fee that optimizes the revenue of the AMM given an optimal arbitrage meaning $\Delta FL^{opt}(f)$. We know that we have to find a fee that fulfils 
\begin{eqnarray}
f<1-\frac{1}{\alpha}\;,
\end{eqnarray}
since otherwise we would suppress the arbitrage attempt.
The revenue $R$ that is made by the AMM is given by 
\begin{eqnarray}
R=\Delta FL^{opt}(\alpha,f)f\;.
\end{eqnarray}
We optimize this expression for a set $\alpha$ which leads to 
\begin{eqnarray}
\frac{dR}{df}\Bigg|_{f=f^{\rm{opt}}}=0\;.
\end{eqnarray}
Solving this equation requires solving for the roots of a third order polynomial. While this is possible in a closed form, it leads to bulky expressions that offer relatively little insight. We are in general interested in cases where $f \ll 1$ and $\alpha \approx 1$. In that case we find an approximate solution given by 
\begin{eqnarray}
f^{\rm{opt}}\approx \sqrt{\alpha}-1\;.
\end{eqnarray}
It is tedious but straightforward to check that under these conditions the revenue of the AMM is twice as high as the revenue of ARB. These implies, that the losses due to arbitrage can be limited to $1/3$ by optimal fee choice, but never fully mitigated or capitalized on. Retaining $2/3$ of the arbitrage activity provides an upper limit for the AMM. 

\subsubsection{Optimal arbitrage vs 'matching the price'}

In the preceding discussion we discussed an arbitrage strategy which maximizes the profit from an arbitrage transaction. What we observed is that this strategy does not match the price of the CEX and the AMM (for here we assume $p_{\rm{CEX}}>p_s$). What it does instead is that it moves the price closer until the ratio hits $\alpha=p_{\rm{CEX}}/p_s=\alpha_{\rm{min}}$. This is when the arbitrage attempt stops buying. The reason for this is intuitive: Imagine trying to match the prices perfectly. Now, instead of doing the transaction in one go, we split the arbitrage operation into two transactions. The first transaction brings $p_s$ closer to $p_{\rm{CEX}}$ but stops at $\alpha_{\rm{min}}$. At this point it wants to make a second transaction doing the remaining alignment. However, as we discussed before, this is a losing operation and consequently will not be completed. A naive arbitrage, however, will align prices and interestingly, the fee dynamics of this is very similar in outcome, but quite different in interpretation.

\section{Simulation results}\label{sec:simulations}

\subsection{Matching the price}

We start with a discussion of the naive arbitrage operation where the prices are simply matched. The condition for profitable arbitrage with the price matching strategy is that $\alpha$ needs to fulfil
\begin{eqnarray}
\alpha>\alpha_{\rm{min}}=\frac{1}{(1-f)^2}
\end{eqnarray}
which can be contrasted from the case of optimal arbitrage where the threshold $\alpha_{\rm{min}}=1/(1-f)$. For practical purposes and realistic fees this implies that the threshold for arbitrage is approximately twice as big for 'matching-the-price' arbitrage as it is for optimal arbitrage meaning the arbitrage frequency has to be lower.

\begin{figure}
\begin{center}
\includegraphics[width=0.8\textwidth]{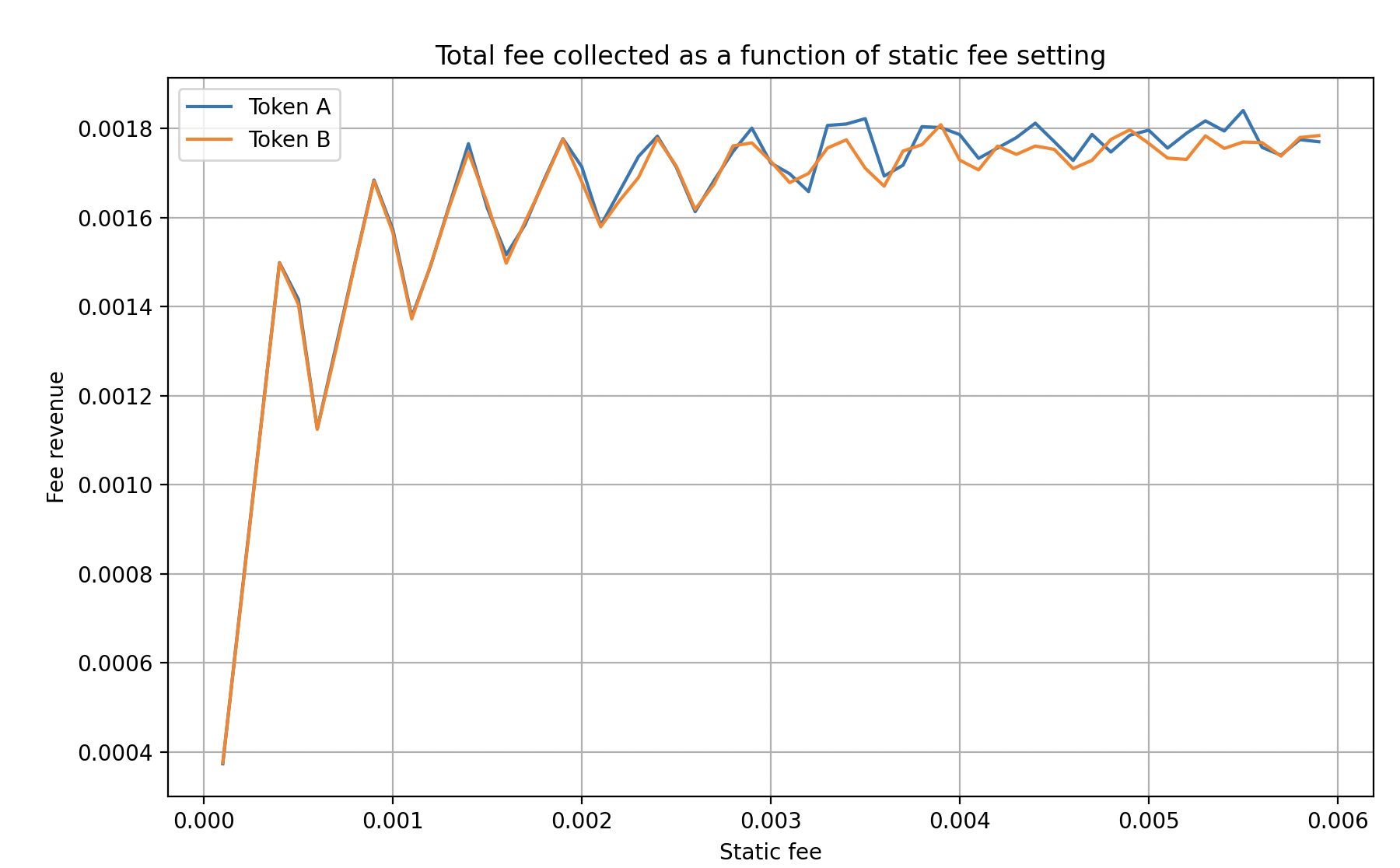}\caption{This plot shows the fee return (y-axis) accrued during a simulation of $1000$ time steps averaged over $1000$ runs. In the plot, blue corresponds to token A and orange to token B. The relative volatility was $0.001$ per time step. The static fees where chosen symmetrically in both directions and varied on the x-axis. The main observation is that fees small compared to the per block volatility lead to suboptimal fee generation (assuming one arbitrage attempt per block) whereas fees on the order three times the per block volatility seem to make a good return.}\label{figure7}
\end{center}
\end{figure}

We have studied the fee revenue creation as a function of the imposed fee. The result of a simulation with $\sigma=0.001$ and $1000$ arbitrage attempts averaged over $1000$ runs is shown in Fig.~\ref{figure7}. The static fees where chosen symmetrically in both directions and varied on the x-axis. The main observation is that fees small compared to the per block volatility lead to suboptimal fee generation (assuming one arbitrage attempt per block) whereas fees on the order three times the per block volatility seem to make a good return. We observe, that there is a saturation for fee settings $f>\sigma$. For fees chosen too high the curve will eventually drop again.

We have checked the frequency with which there are successful arbitrage events and found that it varies $\propto 1/f^2$. This can be explained from the observation of the random walk: For situations with symmetric thresholds we found that the hitting time was $\propto \Delta p^2$. Since $\Delta p_{\rm{arb}}\propto f$ we have that the frequency of hitting should be $\propto f^2$ as confirmed numerically.

The missing ingredients to understanding the observed saturation of the fees is given by
\begin{eqnarray}
R=\Delta FL^{opt}(f)f \propto \sqrt{\alpha_{\rm{min}}}\left( \sqrt{\alpha}-1\right)f \propto f^2
\end{eqnarray}
where the second power in the proportionality stems from $\left( \sqrt{\alpha}-1\right)\propto f$.

Consequently, in the case of no drift, we have $R\propto f^2$ per arbitrage event, while the frequency of successful arbitrage events varies $\propto 1/f^2$ meaning in total we have a constant revenue as function of the fee. 

\begin{figure}
\begin{center}
\includegraphics[width=0.8\textwidth]{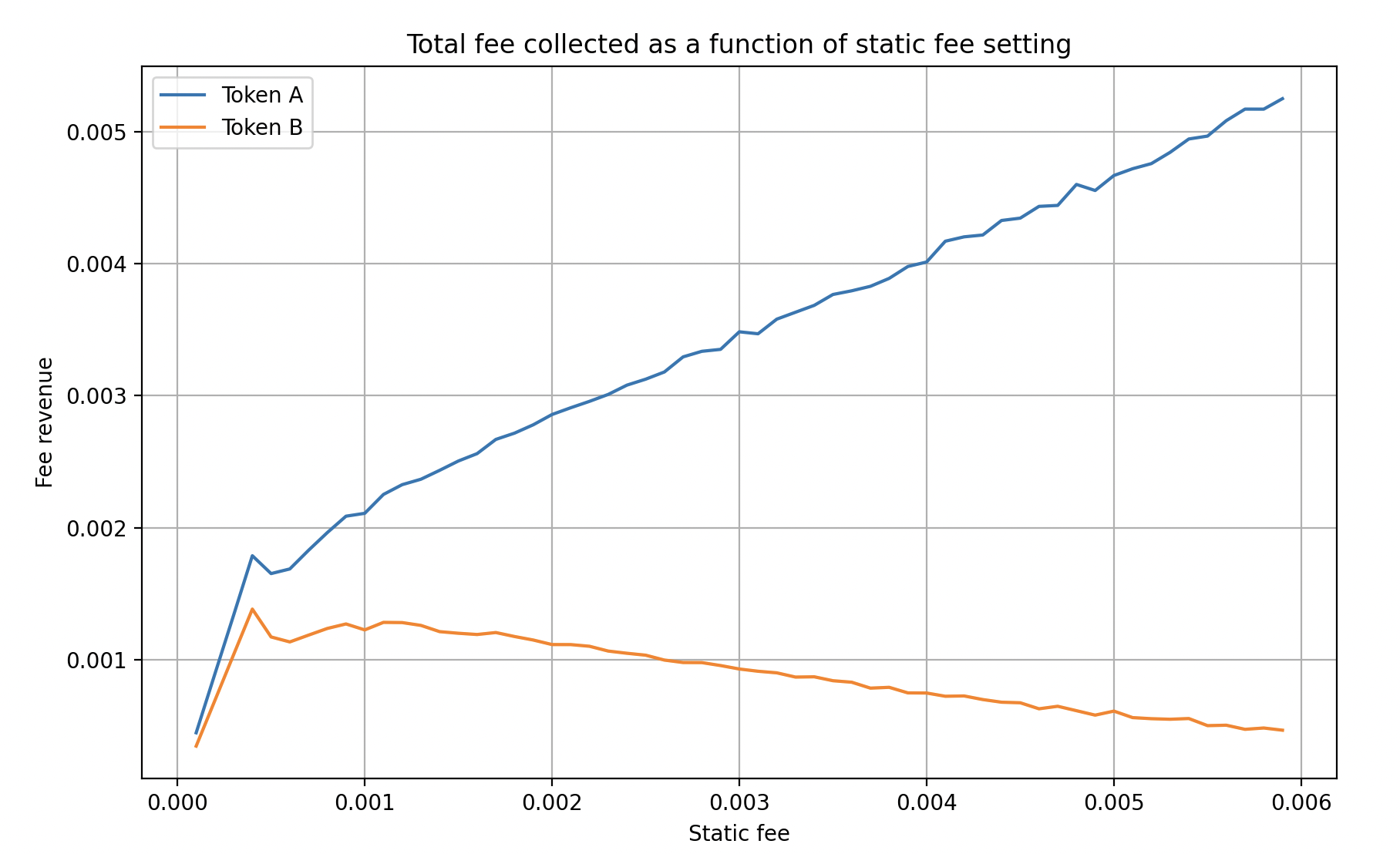}\caption{This plot shows the fee return (y-axis) accrued during a simulation of $1000$ time steps averaged over $1000$ runs. In the plot, blue corresponds to token A and orange to token B. The relative volatility was $0.001$ per time step and we had drift of $\mu=0.0001$. We find linear behavior.}\label{figure8}
\end{center}
\end{figure}

In the case with drift, we still have $R\propto f^2$ but the frequency goes like $1/f$ again following the discussion of the random walk. Consequently, we expect to observe a total revenue that is $\propto f$, in agreement with the observation from the simulation, see Fig.~\ref{figure8}.

\subsection{Optimal arbitrage}

In this part we discuss a perfectly rational arbitrage operation and its interpretation. 

\subsubsection{Volatility without drift}

At first, we make a number of runs in which we fix the volatility at zero drift and simulate a CEX price movement over $10000$ time steps (we can think of these as blocks). After each time step, there is an arbitrage attempt which is declined if it is not profitable. The price on the AMM consequently minimally lags one time step behind the price on the CEX. Our previous analysis also showed that a fully rational ARB does not align the prices perfectly, but only up to a factor $1-f$.

\begin{figure}
\begin{center}
\includegraphics[width=0.8\textwidth]{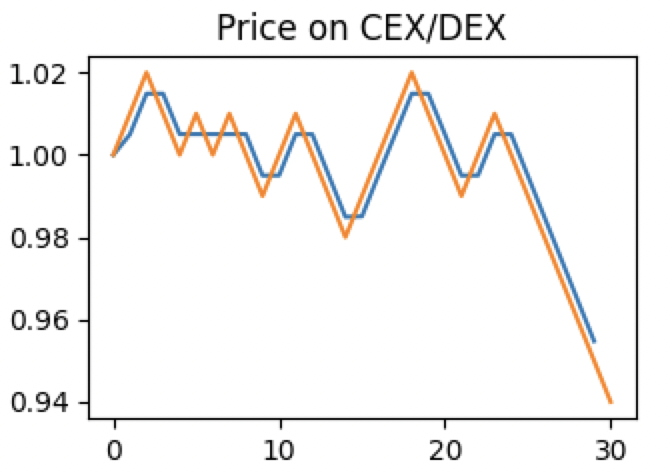}\caption{This plot shows 30 time steps in the price evolution of the CEX (orange line) with a starting price of $1$ and a relative volatility of $0.01$ per time step. The fees where chosen symmetrically in both directions as $0.005$. The AMM contained $15000$ tokens  and $15000$ token B in its initial state. The blue curve shows the price of the AMM as it follows the CEX due to arbitrage. We observe that at some instances, the arbitrage step is skipped as well as that the prices are only aligned up to a factor $1-f$, as discussed previously.}\label{fig2}
\end{center}
\end{figure}

A typical run involving $30$ time steps looks like shown in Fig.~\ref{fig2} with the settings discussed in the figure caption.
An interesting question is whether there is a preferred fee setting that allows to maximize the retention of arbitrage gains. To this end, we make the following experiment: we fix the relative volatility of the CEX price per time step at $0.001$. Consequently, we evolve the CEX price for $1000$ time steps and arbitrage the AMM. In parallel, we collect the fees made during that time. We average this number over $1000$ runs for a series of fees (note that on average the price after the whole series is $1$). 
The expectation is as follows: For smaller fees, one can arbitrage the AMM price more accurately due to more successful attempts. Generically, these attempts are low volume, though, meaning the fees per arbitrage attempt will be moderate. If we choose larger fees, we will have less successful arbitrage events. These, however, will be of higher volume and in addition bring higher fees. Interestingly, we find that overall the concrete choice of fee is irrelevant in the long time limit for sufficiently high fees. 

\begin{figure}
\begin{center}
\includegraphics[width=0.8\textwidth]{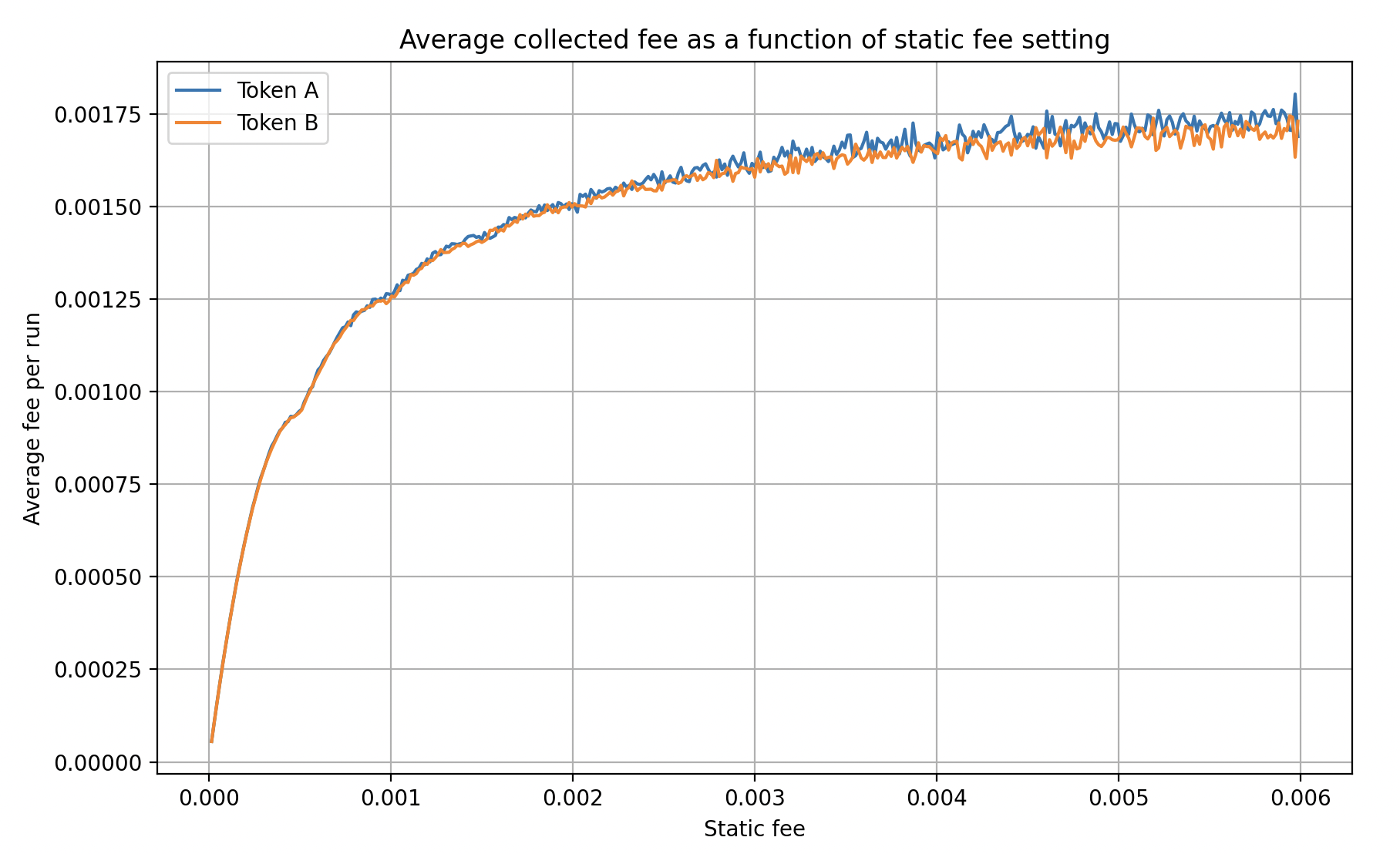}\caption{This plot shows the fee return (y-axis) accrued during a simulation of $1000$ time steps averaged over $1000$ runs. In the plot, blue corresponds to token A and orange to token B. Their agreement correlates with the no-drift setting. The relative volatility was $0.001$ per time step. The static fees where chosen symmetrically in both directions and varied on the x-axis. The main observation is that fees compared small to the per block volatility lead to suboptimal fee generation whereas fees on the order three times the per block volatility seem to make a good return.}\label{fig3}
\end{center}
\end{figure}

\begin{figure}
\begin{center}
\includegraphics[width=0.8\textwidth]{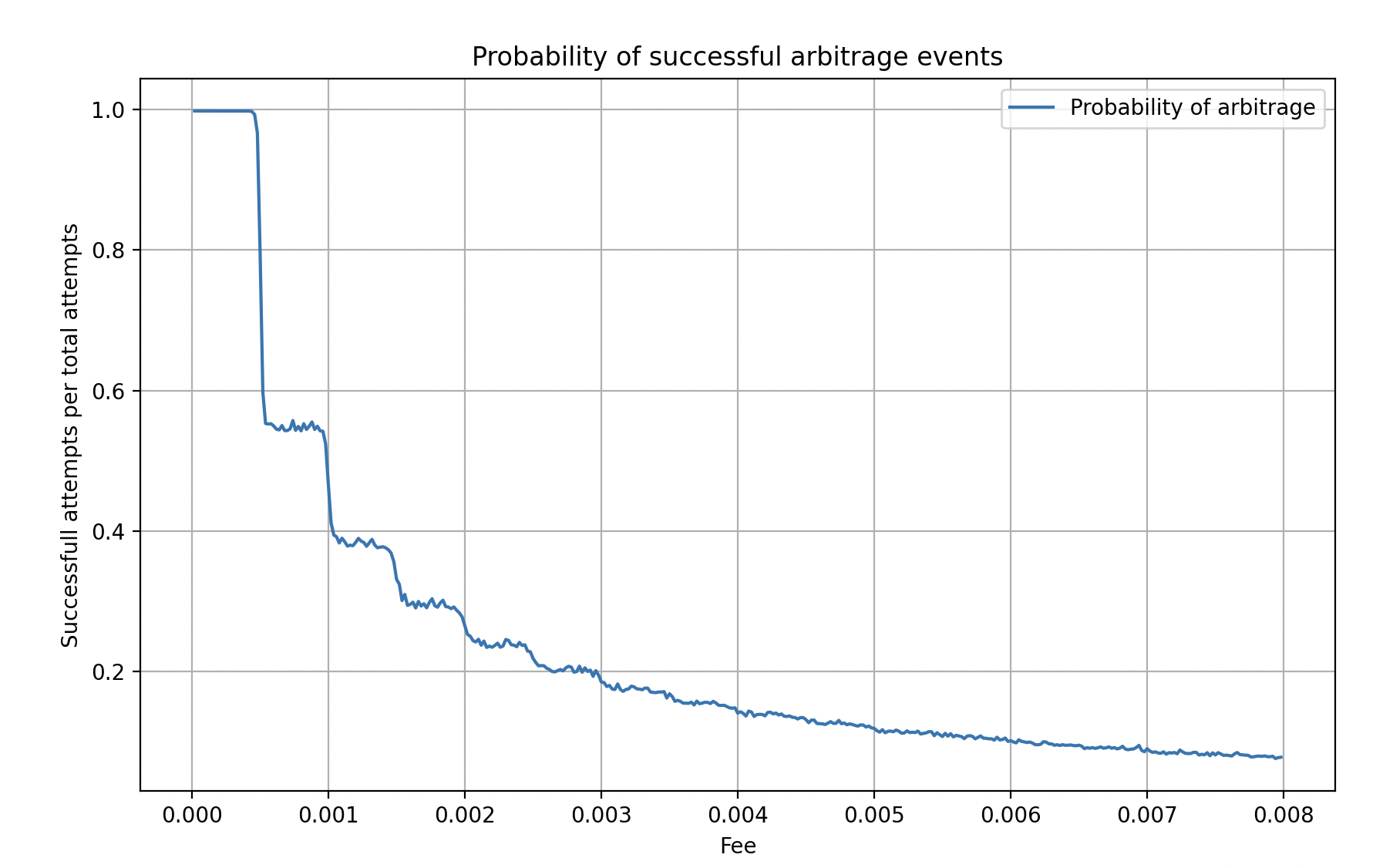}\caption{This plot shows the probability of an arbitrage event to be successful at a given static fee setting. The data was collected in parallel to the data displayed in Fig.~\ref{fig3}.}\label{fig4}
\end{center}
\end{figure}

The corresponding simulation backing this up is shown in Fig.~\ref{fig3}. The y-axis shows the average fees (token A is blue and token B is orange) collected over $1000$ time steps at a given static fee setting, whose value is on the x-axis. In the plot, blue corresponds to token A and orange to token B. Their agreement correlates with the no-drift setting. The relative volatility was $0.001$ per time step. The static fees where chosen symmetrically in both directions and varied on the x-axis. The main observation is that fees compared small to the per block volatility lead to suboptimal fee generation whereas fees on the order three times the per block volatility seem to make a good return.
Additionally, we studied the probability of arbitrage events to be successful (profitable) at a given static fee setting. The result of this is shown in Fig.~\ref{fig4}. The main finding is that for larger fees, the probability of having an arbitrage event follows $\propto 1/f$. This is at odds with the previous section but can be understood following the discussion in Sec.~\ref{sec:technical}. 
The key to understanding is that in contrast to 'matching-the-price' arbitrage, the price is not matched precisely, but only up to the threshold. This implies that in the direction of the last movement, there is effectively no threshold anymore, but any move in that direction triggers arbitrage. In the other direction, however, there is still a minimum price difference requirement. This implies that this corresponds to the case discussed where a random walk has asymmetric thresholds where one threshold is static/fixed (in this case the threshold is effectively zero) and no drift. In that scenario, we found that the hitting time scaled linearly with $\Delta p$ or equivalently with $f$. Consequently, the arbitrage probability goes like $\propto 1/f$, in agreement with the finding. In order to understand the fact that we see saturation it suffices to understand that this allows for arbitrarily small volume swaps that do not scale with the fee so the revenue of the AMM scales linearly with $f$ for every arbitrage trade. Taken together, this explains a constant behavior. 

\begin{tcolorbox}
{\bf Summary of this subsection:}

In a situation with constant volatility and no preferential movement/drift, differences between sufficiently large fee settings average out over time. This can be traced back to the fact that the probability for an arbitrage event decreases $\propto 1/f$ while the gain per arbitrage event grows $\propto f$.  
\end{tcolorbox}

\subsubsection{Volatility with drift}

\begin{figure}
\begin{center}
\includegraphics[width=0.8\textwidth]{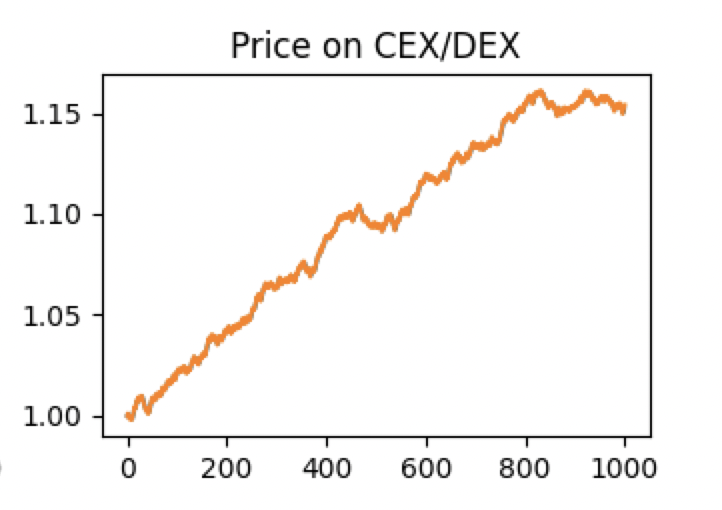}\caption{We simulated $1000$ time steps at relative volatility $0.001$ and drift $0.0001$.}\label{fig5}
\end{center}
\end{figure}

We now move to a situation in which there is a net movement into one direction and study how this changes the whole situation. A representative price curve is shown in Fig.~\ref{fig5} where we simulated $1000$ time steps at relative volatility $0.001$ and drift $0.0001$.

\begin{figure}
\begin{center}
\includegraphics[width=0.8\textwidth]{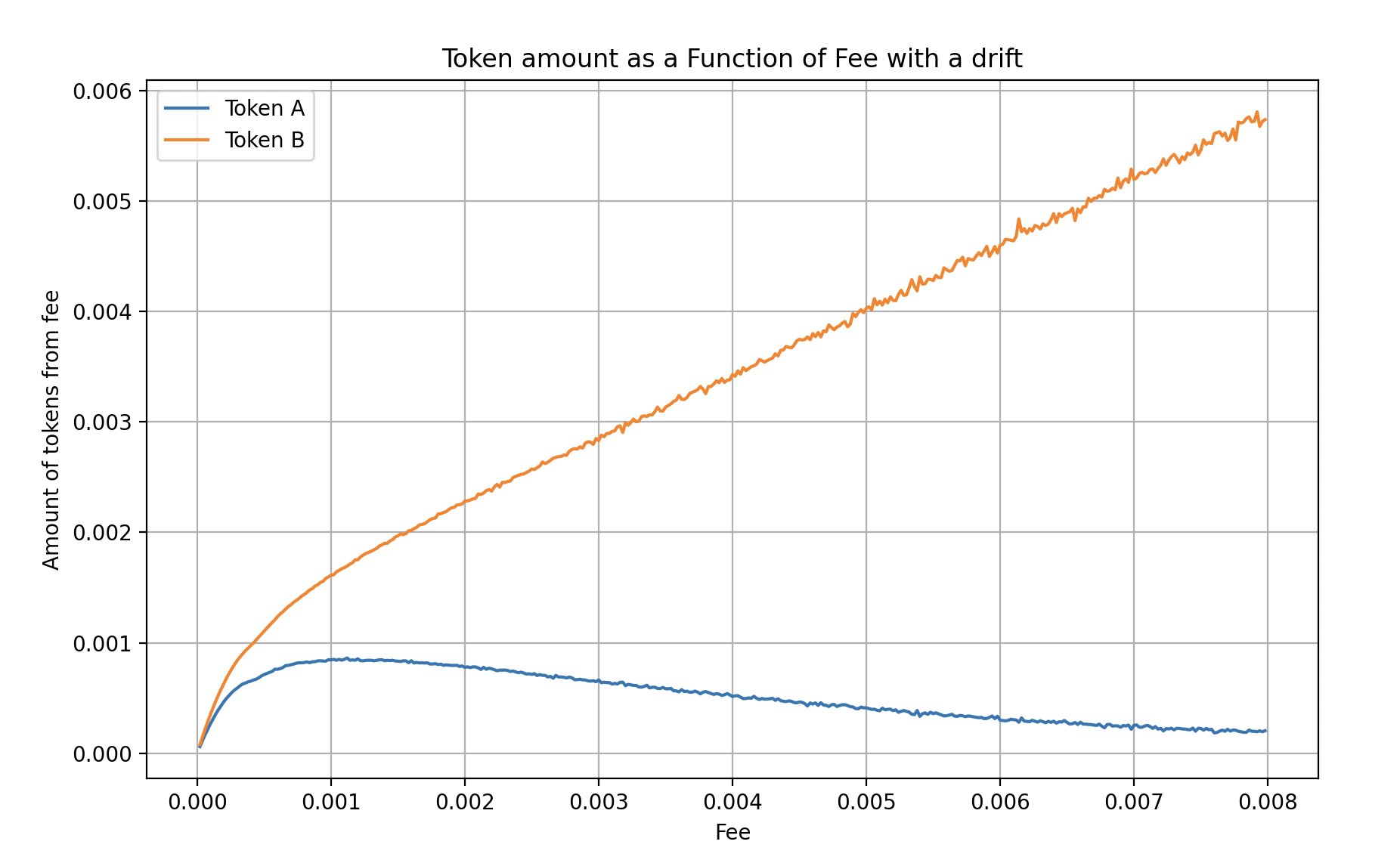}\caption{We simulate $1000$ time steps and average over $1000$ runs. We choose $0.001$ as volatility and $0.0001$ as drift. On average, this leads to a price of $1.1$ after $1000$ time steps.
This plot shows the fee return (y-axis) accrued during a simulation of $1000$ time steps averaged over $1000$ runs. In the plot, blue corresponds to token A and orange to token B. Since the price is going up in the simulation, token B is the predominant fee token. We find no saturation within the window considered meaning we can essentially charge very high fees. For the blue token, however, we find a maximum at the value of the volatility.}\label{fig6}
\end{center}
\end{figure}

We repeat the analysis of the preceding section. We simulate $1000$ time steps and average over $1000$ runs. On average, this leads to a price of $1.1$ after $1000$ time steps.
The plot shows the fee return (y-axis) accrued during a simulation of $1000$ time steps averaged over $1000$ runs. In the plot, blue corresponds to token A and orange to token B. Since the price is going up in the simulation, token B is the predominant fee token. We find no saturation within the window considered meaning we can essentially charge very high fees. For the blue token, however, we find a maximum at the value of the volatility. 
We then analyzed fee generation and found the result shown in Fig.~\ref{fig6}.
This is again simple to understand. We need to recall the results from the case of asymmetric thresholds and drift towards the fixed threshold. There, we found that the average hitting time saturates meaning the associated probability is also constant (we have not added this plot here). Since the revenue per arbitrage still scales linear in $f$, overall we find a linear in $f$ behavior, explaining the numerical findings/
\begin{tcolorbox}
{\bf Summary of this subsection:}

In a situation with constant volatility and  preferential movement/drift, the revenue depends linearly on the fees for the majority token. The minority token saturates and even decays. We find that the probability of having an arbitrage event becomes constant while the revenue per trade grows linearly with the fee $f$, leading to an overall linear behavior. 
   \end{tcolorbox}

\section{Conclusion}\label{sec:conclusion}

In this paper we studied arbitrage dynamics of a simple AMMs. The price discovery was simulated by Brownian motion and constructed to resemble a high liquidity reservoir, like a CEX. The AMM was only subject to toxic flow from the arbitrage activity. We studied how sensitive the revenue generation due to fees is to the actual fee settings. We find that for sideways motion without preferential direction the revenue generation does not depend on the fee setting in the long time limit. However, in the case with drift there is a strong dependence that is directional. One possible usecase for this is to have dynamical fees that react very sensitively to market conditions. Especially, they should pick up directional preferences through asymmetric fees, different for the both tokens in question. It appears like a promising idea to develop fast algorithms that optimize fee settings based on market information allowing for asymmetry in the fees. 

\section*{Acknowledgments}

The authors would like to acknowledge discussions with Fay\c{c}al Drissi and Yuval Boneh.

{\small
  \bibliographystyle{apsrev4-1}
  \bibliography{references}
}

\end{document}